# Origin of spin-driven ferroelectricity and effect of external pressure on the complex magnetism of 6*H*-perovskite Ba$_3$HoRu$_2$O$_9$

E. Kushwaha,[1,#] G. Roy,[1,#] M. Kumar,[1] A. M. dos Santos,[2] S. Ghosh,[1] D. T. Adroja,[3,4] V. Caignaert,[5] O. Perez,[5] A. Pautrat,[5] T. Basu[1, *]

[1]Department of Sciences and Humanities, Rajiv Gandhi Institute of Petroleum Technology, Jais, Amethi, 229304, India
[2]Neutron Scattering Division, Oak Ridge National Lab, Oak Ridge, TN 37831, USA
[3] ISIS Neutron and Muon Source, STFC, Rutherford Appleton Laboratory, Chilton, Oxon OX11 0QX, United Kingdom
[4]Highly Correlated Matter Research Group, Physics Department, University of Johannesburg, Auckland Park 2006, South Africa
[5]CRISMAT Normandie Université, ENSICAEN, UNICAEN, UMR CNRS 6508, 6 Boulevard Maréchal Juin, 14050 Caen Cedex 4, France.

*Corresponding Author: tathamay.basu@rgipt.ac.in

[#]EK and GR have equal scientific contributions to this work.

## Abstract:

The compound Ba$_3$HoRu$_2$O$_9$ magnetically orders at 50 K ($T_{N1}$) followed by another complex magnetic ordering at 10.2 K ($T_{N2}$). The 2$^{nd}$ magnetic phase transition was characterized by the co-existence of two competing magnetic ground states associated with two different magnetic wave vectors ($k_1$= ½ 0 0 and $k_2$=¼ ¼ 0). The multiferroicity and magnetoelectric coupling were predicted below $T_{N2}$ in these 4*d*-based materials. Here, we have discussed the origin of spin-driven ferroelectricity, which is not known yet, and the nature of magnetoelectric domains. We have investigated the compound through time-of-flight neutron diffraction, synchrotron X-ray diffraction (XRD), AC susceptibility, frequency-dependent complex dielectric spectroscopy, and DC magnetization under external pressure. We demonstrate that the non-collinear structure involving two different magnetic ions, Ru(4*d*) and Ho(4*f*), break the spatial inversion symmetry via inverse Dzyaloshinskii–Moriya (D-M) interaction through strong 4*d*-4*f* magnetic correlation, which shifts the oxygen atoms and results in non-zero polarization. Such an observation of inverse D-M interaction from two different magnetic ions which caused ferroelectricity is rarely observed. The stronger spin-orbit coupling of 4*d*-orbital might play a major role in creating D-M interaction of non-collinear spins. We have systematically studied the spin and dipolar dynamics, which exhibit intriguing behavior with shorter coherence lengths of 2$^{nd}$ magnetic phase associated with the k$_2$-wave vector. The results manifest the development of finite-size magnetoelectric domains instead of true long-range ordering which justifies the experimentally obtained low value of ferroelectric polarization. The lattice parameters and volume show a sharp anomaly at T$_{N2}$ obtained by analyzing the temperature dependence XRD which is consistent with the ferroelectric transition, predicting a non-centrosymmetric space group, P$\bar{6}$2c, for this compound. Furthermore, we have investigated the effect of external pressure on this complex magnetism. The result reveals an enhancement of ordering temperature by the application of external pressure (~1.6 K/GPa). The external pressure might favor stabilizing the magnetic ground state associated with 2$^{nd}$ magnetic phase. Our study shows an unconventional mechanism of spin-driven ferroelectricity involving inverse D-M interaction between Ru(4$d$) and Ho(4$f$) magnetic ions due to strong 4$d$-4$f$ cross-coupling.

## 1. Introduction:

The multiferroic materials, in which, magnetic and ferroelectric phases coexist, and where spins and electric dipoles are strongly coupled, have picked up momentum in condensed matter physics in recent years due to their future potential application in memory devices and fundamental interest.[1–3] Several systems with 3$d$-4$f$ coupling that show multiferroic behavior are known with many of these widely studied: e.g, RMnO$_3$, [1,2] RMn$_2$O$_5$, [3–5] RCrTiO$_5$, [6,7] R$_2$BaNiO$_5$ [8,9] where R is a rare earth. However, in these systems, spatial inversion symmetry breaks due to complex patterns associated with the 3$d$ transition metal ion. This arises from the coupling of strong electronic correlations with magnetic frustration, [2,10,11] specifically the antisymmetric Dzyaloshinskii–Moriya (DM) interaction,[12,13] wherein non-collinear spins break spatial inversion symmetry, inducing a non-zero polarization in materials; [1] Exchange-striction, at which lattice distortion is induced by the symmetric exchange interaction between neighboring magnetic ions in a material, leading to changes in the crystal structure, [14] and spin–ligand interaction (spin-dependent p–d hybridization) where the locally polar bond connecting the spin site and the ligand site can be modulated by the spin-direction-dependent hybridization arising from the spin-orbit coupling [2], among other factors. These multiferroic systems exhibit fascinating multiferroic phenomena and magnetoelectric (ME) domain dynamics in which the specific spin pattern of 3$d$-magnetic ion is mainly responsible for creating electric dipole moments (i.e. non-zero polarization). [2,8,15–18] For example, the electric polarization in well-known multiferroic compounds, such as $RMnO_3$, $RMn_2O_5$, $RCrTiO_5$, $R_2BaNiO_5$, $Ba_3NiNb_2O_9$, $CuCrO_2$, $Co_4Nb_2O_9$, $Ca_3CoMnO_6$, arises from canted spin-structure of $3d$-magnetic ion (Co, Mn, Ni, Cr, etc.) due to inverse D-M interaction mechanism or from collinear spin-structure $3d$-magnetic ion due to the exchange-striction mechanism, or due to p-d hybridization. [1,2,6,11,14] The rare-earth (R)-ions might have a key role in the resulting overall magnetic structure of $3d$-transition metal ions on these systems, but spins of R-ions are not directly involved in the mechanism of spin-driven ferroelectricity. For example, in the well-known $RMnO_3$ compound, the non-collinear structure of Mn gives rise to polarization due to inverse D-M interaction, where R-ion has an indirect role in the Mn-spin pattern. [1,2] However, the spin-driven polarization arising from inverse D-M interaction between two different magnetic species has rarely been reported. A little theoretical study exists on this aspect [9,19,20], but experimental reports are rare in this aspect. For the magnetoelectric compound $NdCrTiO_5$, theoretical studies suggest that the spin-driven polarization emerges due to inverse D-M interaction between non-collinear spins of both Cr and Nd, [20] however there is no experimental evidence of this spin-structure. The significant role of both Tb and Ni spins (non-collinear spins) on magnetoelectric coupling is also predicted theoretically for the multiferroic compound $Tb_2BaNiO_5$ though no clear mechanism is demonstrated. [9] For the well-known multiferroic series $RMn_2O_5$, the compound $GdMn_2O_5$ shows much larger polarization compared to other R-members. [4,5] It is described that in addition to the Mn-Mn exchange-striction, the exchange-striction of Gd-Mn (co-linear spins) contributes to the net ferroelectric polarization yielding a larger polarization in this compound when compared with the remaining members of the family. [4,19] No other study exists on this aspect where both spins of transition-metal ($d$) and rare-earth($f$) directly take part in breaking the inversion symmetry. Moreover, all these studies are restricted to $3d$-transition metal. [9,19–22] The search for potential multiferroic material in the hybrid metal-organic network has been explored as well, where magnetodielectric coupling is documented but spin-driven ferroelectricity is not demonstrated. [23–25] Recently, the search for potential multiferroic materials was extended to strongly correlated $4d/5d$-orbital systems to investigate the nature of ME coupling. The pronounced effect of stronger spin-orbit coupling, the larger radial extension of $d$-orbital, and the crystal field effect in $4d/5d$-orbital, when compared to that of 3d-orbital, is considered crucial in magnetism, multiferroicity, and thus ME coupling. For example, a stronger spin-orbit coupling may help to enhance the ME coupling strength through inverse D-M interaction, whereas the crystal field could have a significant effect in deciding the magnetic ground state as well as lattice distortion. However, the larger extension of the $4d/5d$ orbitals, when compared to that of $3d$, makes the sample less insulating and hinders the experimental determination of the polarization. Therefore, despite strong theoretical prediction, [26] experimental reports of $4d/5d$ (non-$d^0$)-based multiferroic (spin-driven ferroelectric) compounds are rare. Very recently, magnetodielectric coupling and multiferroicity were experimentally demonstrated in a $4d$-$4f$ coupled system, $Ba_3HoRu_2O_9$, [27–29] which has increased research activities in this field for searching for $4d$-transition metal oxide-based multiferroic systems. Further investigations are warranted to understand the mechanism of the (magnetism-driven) ferroelectricity, which has not yet been explored.

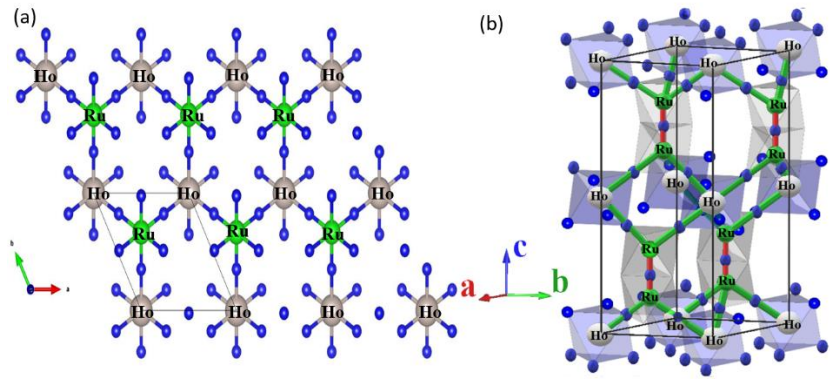

**Figure 1:** Crystal structure of $Ba_3HoRu_2O_9$ (Green (Ru), Grey-white (Ho), and Blue (Oxygen), Ba-atoms are removed for a clear view to show superexchange path. (a) in ab-plane, and (b) 3D view to show the $HoO_6$ and $RuO_6$ octahedra and Ru-O-Ru superexchange path (in red) and Ho-O-Ru superexchange path (in green).

Another interesting area is the effect of external parameters (e.g., pressure, magnetic field) on multiferroicity. These have a huge influence on spin and dipoles resulting in distortion in lattice and change in superexchange magnetic interaction which varies with rare earth metal, even in the same series due to competing magnetic interactions. For example, an interesting re-entrant multiglass behavior below the multiferroic ordering temperature is observed in the Haldane-chain system $R_2BaNiO_5$, in which the external magnetic field has an intriguing effect. [15,16,30] The magnetic and dielectric ordering both shifts to lower temperature under the application of magnetic field for the compound $Er_2BaNiO_5$, [15] whereas, the dielectric peak shifts in the opposite direction compared to that of magnetic feature in-presence of magnetic field for the isostructural $Dy_2BaNiO_5$ -an observation that still requires further investigations. [16] A large enhancement of pressure-induced multiferroicity (ordering temperature or/and value of electric polarization) by stabilizing the magnetic phase-competing ground state has been theoretically predicted in many multiferroic systems (e.g., CuO, PbCoO$_3$) [29,30]. However, the corresponding experimental realization is very limited (e.g., GdMn$_2$O$_5$, PrMn$_2$O$_5$, Mn$_2$GeO$_4$). [19,33,34]

The compound that is at the center of this report, Ba$_3$HoRu$_2$O$_9$, crystallizes in a *6H*-perovskite structure and consists of Ru$_2$O$_9$ dimers (face-sharing distorted RuO$_6$ octahedral) and regular corner-sharing RO$_6$ octahedral (Figure 1). [29,35] This system undergoes an antiferromagnetic ordering at T$_{N2}$~ 10.2 K. The magnetic super-exchange interaction develops through the path "Ru-O-Ho-O-Ru" (Figure 1). It is demonstrated that magnetodielectric (MD) coupling is stronger for heavy rare-earth members. [29] Among all the known compositions, Ba$_3$HoRu$_2$O$_9$ exhibits the strongest MD coupling and shows ferroelectricity with a weak polarization value below the magnetic ordering (T$_{N2}$=10.2 K). [29] Our recent neutron diffraction studies [27] reveal that the compound Ba$_3$HoRu$_2$O$_9$ exhibits magnetic ordering at T$_{N1}$=50 K, associated with a propagation wave vector of k$_1$ = (0.5 0 0), followed by another magnetic phase transition around T$_{N2}$=10.2 K that results in a more complex magnetic phase. In contrast, all other members in this family studied (R= Tb, Gd, Er, Sm, etc.) till now, show only the low-temperature antiferromagnetic ordering (~10-12 K), [27,29,35] except Ba$_3$NdRu$_2$O$_9$ which shows ferromagnetic ordering at 24 K (where Nd-moments orders) followed by another ordering at 17 K (Ru-moments also order) where Ru$_2$O$_9$-dimers order antiferromagnetically. [36] Interestingly, for our titled compound Ba$_3$HoRu$_2$O$_9$, both Ru and Ho-moments order simultaneously below T$_{N1}$, followed by spin reorientations at lower temperatures, revealing a strong Ru(4$d$)-Ho(4$f$) magnetic coupling in this system. [27] Below T$_{N2}$, in addition to a sharp reorientation of Ho and Ru spin-moments, another magnetic phase with a propagation wave vector k$_2$ = (¼ ¼ 0) emerges and coexists with the one associated with k$_1$, revealing a competing magnetic ground state. [27] The 2$^{nd}$ magnetic phase is characterized by an up-up-down-down (↑↑↓↓) spin-structure of each Ru and Ho-spins. The compound exhibits a peak in the T-dependent complex dielectric constant at the onset of the 2$^{nd}$ magnetic phase transition (T$_{N2}$) which has a pronounced effect in the presence of a magnetic field, mimicking the magnetic feature observed in magnetic susceptibility and heat capacity. [29] This confirms magnetodielectric coupling. Further, positive-up-negative-down (PUND) ferroelectric polarization measurements at 5 K (below T$_{N2}$) confirm the ferroelectricity characterizing this compound as multiferroic. [29]

However, it is not clear how spin drives the ferroelectric polarization. The studies presented here contribute to the understanding of the nature of magnetic ordering and spin dynamics around $T_{N2}$ and, consequently, elucidate the mechanism of magnetoelectric coupling and multiferroicity in this compound. We demonstrate that inverse D-M interaction between Ru(4*d*) and Ho(4*f*) spins related to the spin-structure of the 2$^{nd}$ magnetic phase transition breaks the inversion symmetry producing non-zero electric polarization. In addition, we have studied the effect of external pressure. Our results suggest that external pressure enhances the magnetic ordering, probably by reducing the magnetic phase competition.

## 2. Experimental Details:

The powder sample $Ba_3HoRu_2O_9$ was synthesized by solid-state-reaction using mixtures of high purity (>99.9%) precursors: $BaCO_3$, $RuO_2$, and $Ho_2O_3$ which were mixed in an agate mortar and pestle and pressed into pellets as described in Ref. [29,35] The compound forms a single phase with $P6_3/mmc$ space group as reported earlier. [29,35]

Temperature-dependent AC magnetic susceptibility measurements were carried out using a Quantum Design Superconducting Quantum Interference Device (SQUID). The high-pressure SQUID measurement was done using a beryllium copper homemade clamp-style cell in Oak Ridge National Laboratory (ORNL). The sample magnetization is calculated by subtracting the empty pressure cell data. The ambient pressure (standard sample loading) dc magnetization is conducted in the presence of 100 Oe-50 kOe magnetic field in zero-field-cooled (ZFC) mode (the sample is cooled under zero field from paramagnetic region to 2K and then the magnetic field is applied, the data has been taken during warming) and after that, the dc magnetization is taken under 10 kOe magnetic field applying different pressure (0-1.2 GPa) in the same ZFC mode.

The complex dielectric measurements as a function of temperature, magnetic field, and frequency (1 V AC bias, 1-100 kHz) were performed using a LCR meter (Agilent 4284A). This measurement set-up is integrated to a Quantum Design Physical Properties Measurement System (PPMS). Silver paint was used to make parallel-plate capacitors of the pressed disc-like polycrystalline samples.

Neutron Powder diffraction patterns were collected at the SNAP beamline, a time-of-flight diffractometer dedicated to high-pressure research at Spallation Neutron Source (SNS), Oak Ridge

Laboratory, USA. To cover the regions of interest in reciprocal space the detector banks were placed at a central scattering angle of 50º and 65º about the sample in the scattering plane (each spanning a 45º range) and two central wavelengths selected were 2.4 Å and 6.4 Å, leading to an incident beam with usable wavelength spectra of 0.65 Å to 4.15 Å, and 4.65 Å to 8.15 Å (second frame). The resulting diffraction data covered a range, in momentum transfer $q$ from 0.9 Å$^{-1}$ -10 Å$^{-1}$ or, in d spacing from 0.5 Å < d < 7 Å.  In the second frame, the corresponding ranges are momentum transfer $q$ from 0.4 Å$^{-1}$ – 1.8 Å$^{-1}$ or, in d spacing from 3.5 Å < d < 12 Å. The available low-Q range in time-of-flight neutron diffraction measurement will be able to track any other magnetic peak (if appears) for both propagation vectors $k_1$ and $k_2$.

The sample was placed in a liquid He cooled wet "orange" cryostat and data were collected from room temperature down to 4 K. Two sets of measurements were attempted: one at ambient pressure where the sample was loaded on a thin walled 6 mm diameter vanadium can and another set using a high strength CuBe clamp pressure cell with 10 mm in diameter.

Initial analysis indicated that the elevated background levels of the relatively smaller amounts of sample measured and the weak magnetic scattering in the pressure cell precluded the tracking of the magnetic phases with pressure. As such in the following, only the ambient pressure data is presented.

We have carried out the temperature dependence high-resolution X-ray diffraction (HRXRD) in CRISTAL beamline in SOLEIL synchrotron (France) with an X-ray source of wavelength $\lambda$= 0.58182 Å.

## 3. Results and Discussions:

### *3.1. Time-of-flight Neutron diffraction and Mechanism of Spin-driven ferroelectricity:*

Here, we will elucidate the mechanism of spin-driven polarization. To confirm the magnetic structure, we have performed neutron diffraction at SNAP (ORNL), which allow us to measure the low-Q regime to detect all the magnetic peaks of both magnetic phases associated with $k_1$ (½ 0 0) and $k_2$ (¼ ¼ 0), as shown in Figure 2a. We have reproduced the same spin-structure as obtained in our earlier report. [27] The spin structure shown in Figures 3a and 3b are associated with $k_1$ and $k_2$ wave vectors at 4 K.

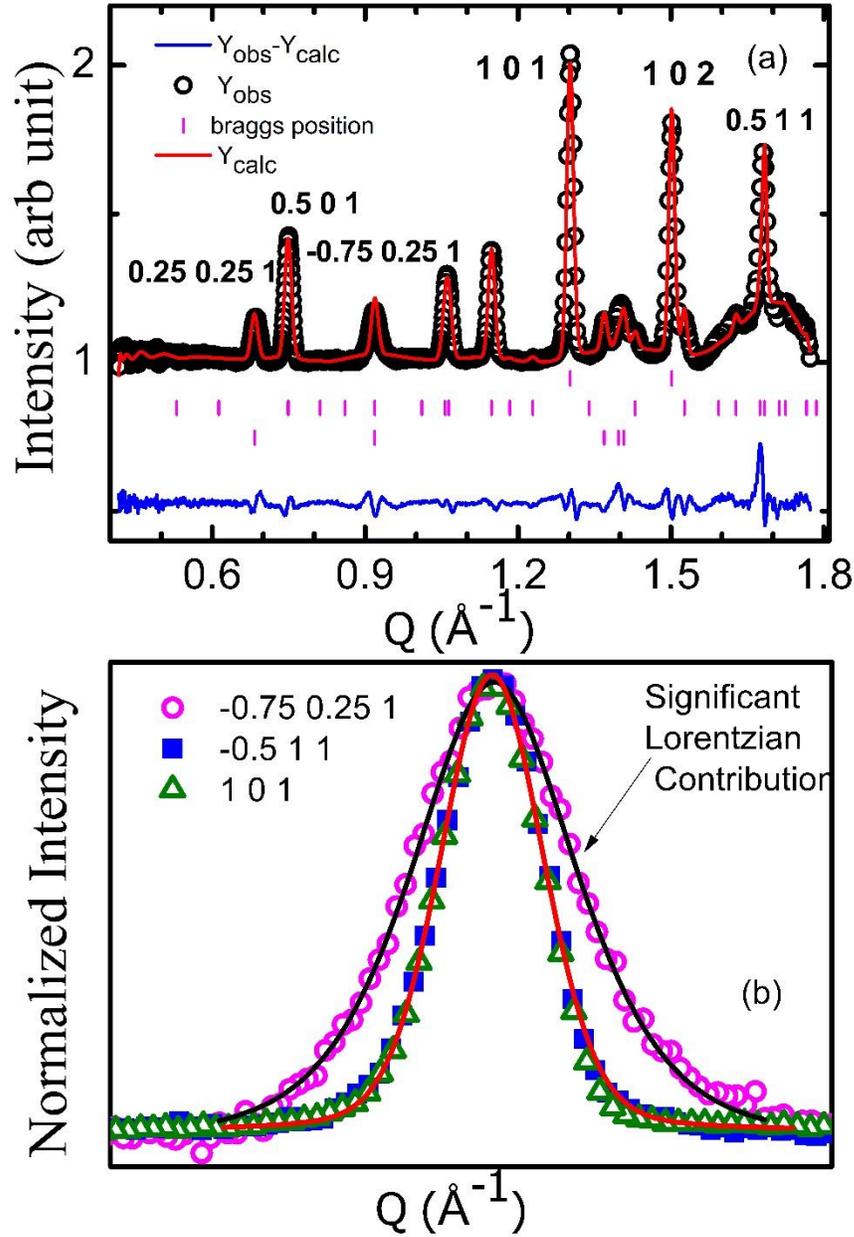

**Figure 2:** (a) Rietveld Refinement of Time-of-flight neutron diffraction data in the low Q regime at T=4 K of $Ba_3HoRu_2O_9$. The open black circle represents the experimental data, while the red solid line shows the Rietveld fitting. The vertical bars display the Bragg peak positions: The upper vertical lines are Bragg lines for the crystal structure, followed by reflections associated with $k_1$ = (0.5 0 0), and $k_2$= (0.25 0.25 0) respectively. The lower blue line is the difference between the experimental and calculated intensity (b) Normalized peaks for Lattice (1 0 1), $k_1$ (-0.5 1 1) & $k_2$ (-0.75 0.25 1) magnetic structure as discussed in the Text. The value of the x-axis (Q) is shifted for each plot to place the peak at the same position to compare the peak shape directly. The solid line is fitting of these peaks. The solid "red" line is fitted with Gaussian and the solid "black" line is fitted with Gaussian and Lorentzian parameters.

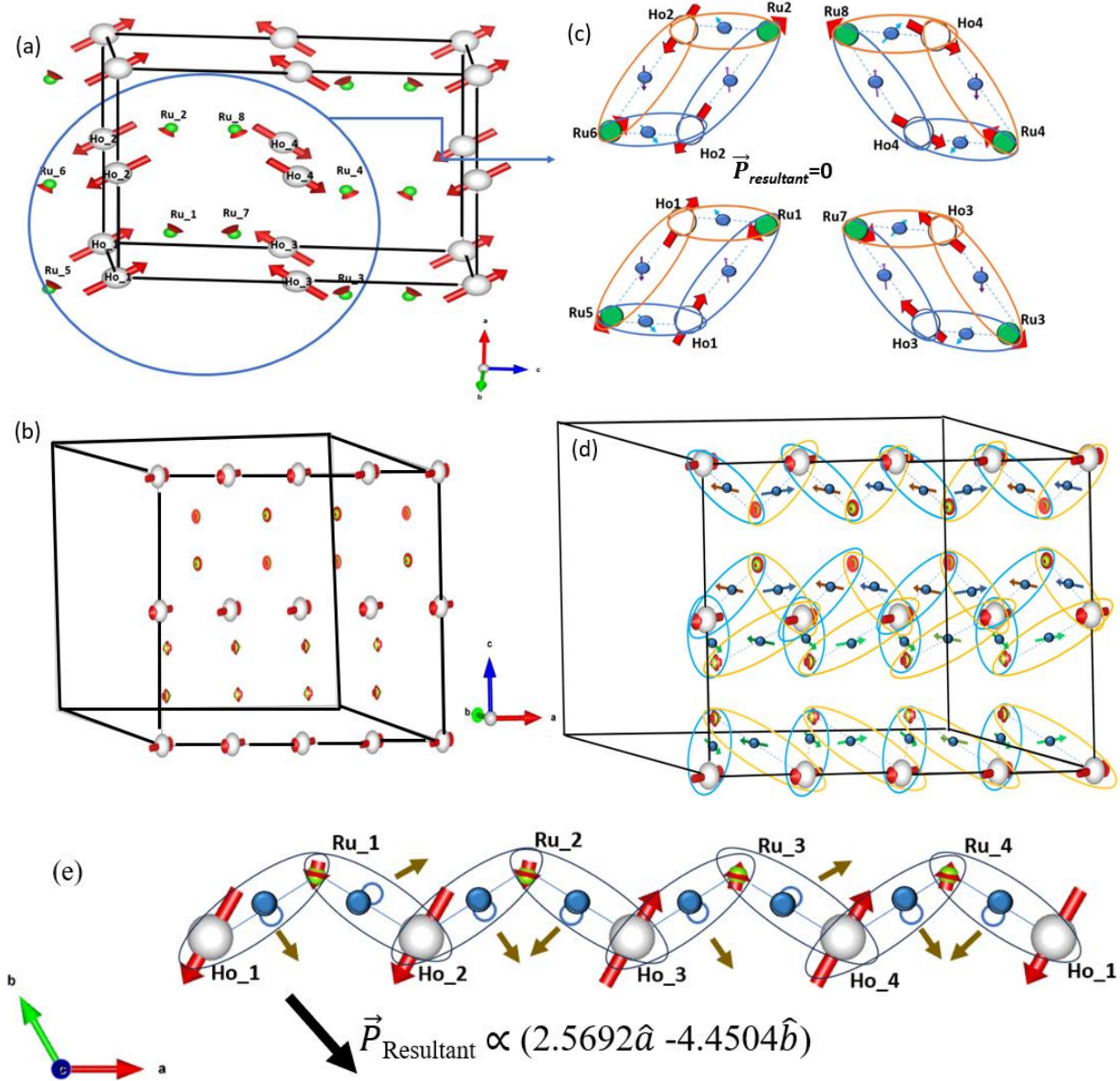

**Figure 3:** Magnetic structure of (a) 1st magnetic phase (canted structure) associated with k$_1$-wave vector, and (b) 2nd magnetic phase (up-up-down-down structure) associated with k$_2$-wave vector at *T*=4 K for Ba$_3$HoRu$_2$O$_9$. The oxygen atoms in between Ru and Ho atoms are not shown in the figure.

(c) and (d) calculated inverse D-M interaction considering all neighboring Ru and Ho spin for each structure respectively, as discussed in the text. The atoms are denoted in a different color as follows: White (Ho), Green (Ru), and Blue (Oxygen). The red arrow on Ho and Ru -atoms denotes the spins of the respective atoms. The arrow on oxygen shows the direction of local polarization, that is, the direction of inverse DM interaction between related Ho and Ru spins of corresponding Ho-O-Ru configuration. The large black arrow shows the direction of resultant polarization. The resultant polarization is zero for (c) and non-zero for (d).

(e) The spin-structure and displacement of oxygen atoms due to inverse D-M interaction between Ho and Ru spins are schematically shown. The displacement of oxygen atoms (closed blue circle) is schematically depicted with open blue circle, which is along the direction of inverse D-M interaction (shown with a brown arrow). The resultant polarization (that is proportional to inverse D-M interaction) does not cancel out in the macroscopic scale, shown with a large black arrow. The detailed calculation is described in **S.I.** [37]

The electric polarization in this system most likely may arise through inverse Dzyaloshinskii-Moriya interactions from non-collinear magnetic structure (spin-current model) or through exchange-striction from the collinear spin-structure. First, we have calculated inverse D-M interaction [$P_{ij} \propto e_{ij} \times (S_i \times S_j)$] between nearest Ru and Ho-spins (Ru-O-Ho configurations) of magnetic phase related to $k_1$-spin structure, shown in Figure 3c for T= 4 K. This results in local polarization, however, the direction of the polarization of one Ru-O-Ho configuration is opposite to the neighboring configuration. Hence, the net polarization is cancelled out in the whole magnetic unit cell (in macroscopic scale), as shown in Figure 3c. We have also calculated the same at T=30 K (not shown here), which does not yield any macroscopic polarization as well. Therefore, the inverse D-M interaction associated with non-collinear $k_1$-spin structure cannot produce ferroelectric polarization. The system magnetically orders below 50 K ($T_{N1}$) associated with $k_1$-spin structure. The absence of ferroelectricity in $T_{N1}<T< T_{N2}$ agrees with the above conclusion. The clear peak in dielectric constant is observed at $T_{N2}$ and ferroelectricity is observed below $T_{N2}$, where 2$^{nd}$ magnetic phase related to $k_2$-spin structure emerges. Initially, it was assumed that the spin-driven ferroelectricity might arise from exchange-striction of ↑↑↓↓ spin-structure of Ho or Ru from 2$^{nd}$-magentic phase, similar to the observed in the $Ca_3CoMnO_6$ system. [14] In the exchange-striction mechanism, the symmetric exchange interaction between the parallel spins (↑↑ or ↓↓) helps to make a shorter bond which moves the ion closer, subsequently, opposite (antiparallel) spins (↑↓) make the bond longer between these two ions, which creates a local polarization. Here, the exchange-interaction is mediated via oxygen ions. For ↑↑↓↓-spin structure ($k_2$) in ab-plane of this compound, the Ho ions are connected to next Ho-ions via Ho-O-Ru-O-Ho exchange-path or via Ho-O-Ba-O-Ho path (see Figure 1), there is no Ho-O-Ho path exists. Thus, the superexchange interaction between two nearest Ho-spins is not possible, the superexchange path should be via Ru (see Fig.S1 in Supporting Information (S.I.) [37]). Hence, ↑↑↓↓ Ho-spins cannot have such exchange-striction effect (negligible displacement) and such a configuration cannot produce any polarization. The Ru-O-Ru super-exchange interaction is restricted in $Ru_2O_9$-dimers, as they are not directly interconnected (a clear view is shown in Fig. S1 in S.I.[37]). The magnetic ordering is obtained via Ru-O-Ho exchange path only. Therefore, the Exchange-striction mechanism in ↑↑↓↓ spin-structure of only Ho or Ru-ions cannot produce dipole moment. In ↑↑↓↓ ($k_2$)- spin-structure, the individual Ru-spins moments or individual Ho-spins moments are collinear in the ab-plane among themselves, nevertheless, the Ho and Ru spins are non-collinear to each other (connected via Ru-O-Ho-O-Ru exchange path) even in ab-plane (see Fig. S1 in S.I.[37]). The Ru-O-Ho-O-Ru superexchange paths are shown in Figure 1 and Figure 3d where Ho and Ru-ions are arranged in a zig-zag pattern. The presence of larger $d$-orbitals ($4d$) results in strong spin-orbit coupling in this system. Normally, stronger spin-orbit coupling may favor the D-M interaction, which may lead to a preference for non-collinear spin configurations of Ho and Ru via strong $d$-$f$ coupling. In this study, we observe a low magnetic moment of the Ru atom, which is 1.1 µ$_B$ compared to its average spin-only values in Ru$^{4+}$ ($d^4$) and Ru$^{5+}$ ($d^3$) states. This observation further suggests the possibility of electron itinerancy with a larger extension of $4d$-orbital, or/and competing effects of crystal field effects and spin-orbit coupling of the Ru atom in this system. We have calculated the inverse D-M interaction between the nearest Ru and Ho (connected via Ru-O-Ho path) of this (k$_2$)- spin-structure, which produces a non-zero value which shifts the oxygen atoms and thus, non-zero local polarization (see Figure 3d and 3e). We have calculated for the whole unit cell. We obtained that the local polarization does not cancel out in the whole unit cell and a non-zero resultant polarization is obtained in ab-plane (Figure 3d and 3e). The detailed spin-structure in different orientations (Fig. S2) and calculation of inverse D-M interaction $[\boldsymbol{P_{ij}} \propto \boldsymbol{e_{ij}} \times (\boldsymbol{S_i} \times \boldsymbol{S_j})]$ is clearly documented in S.I. [37]

To date, there are mostly experimental reports of those multiferroic compounds where spin-pattern of $3d$-metal ion is responsible for spin-driven ferroelectricity, as discussed in the introduction. However, the experimental evidence of electric polarization as a result of inverse D-M interaction of non-collinear spin-moments occurring between two different atoms does not exist (to the best of our knowledge). The multiferroic GdMn$_2$O$_5$ compound exhibits the largest polarization among all multiferroic-II compounds (that is, multiferroic magnetoelectric system where spin-drives the ferroelectricity). It is demonstrated that the exchange-striction mechanism between collinear spins of Gd($4f$) and Mn($3d$) contributed to a larger polarization of GdMn$_2$O$_5$ compound. [19] Here we observe that the spin-driven polarization is governed from non-collinear spin-structure of two different atoms, that is, Ho($4f$) and Ru($4d$) through inverse D-M interaction. This further supports strong $4d$-$4f$ correlation in this system. The stronger spin-orbit coupling in larger $4d$-orbital compared to that of $3d$-orbital might play a decisive role.

## 3.2. Magnetoelectric domain dynamics:

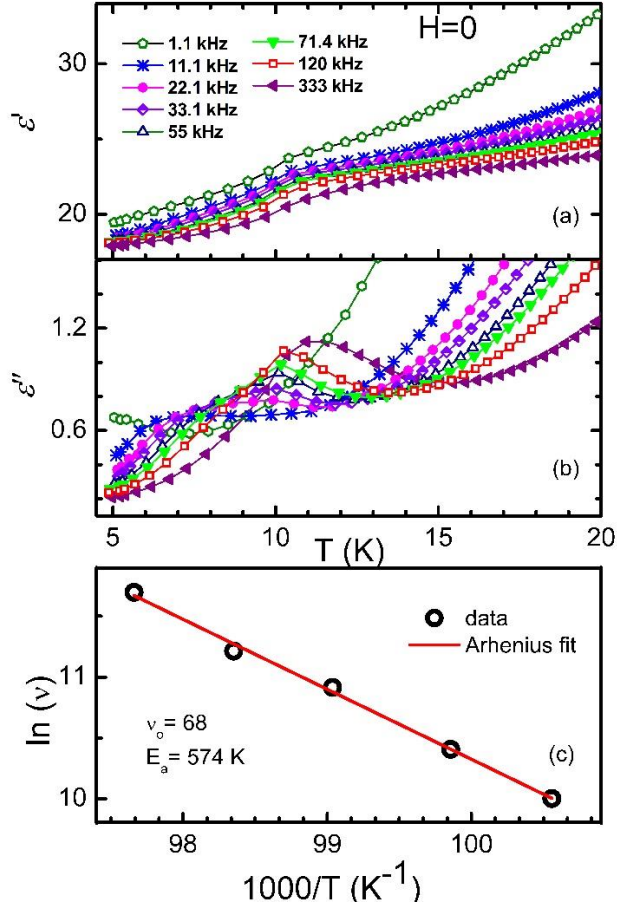

**Figure 4: (a) The real and (b) imaginary part of dielectric constant as a function of temperature under various frequencies (1.1 -120 kHz). (c) Arrhenius plot ln$v$ versus inverse of $T_p$. The low frequencies data in (c) is not included either due to absence of the corresponding peak or very broad features.**

The polarization through positive-up-negative-down (PUND) measurement was reported to be very low. [29] To understand this behavior, we have investigated the spin and dipolar domain dynamics of the $Ba_3HoRu_2O_9$ compound through combined dielectric characterization, ac magnetic susceptibility, and neutron time-of-flight measurements. We have measured the dielectric constant as a function of temperature in the absence magnetic field at various frequencies from $v$= 1.1-120 kHz, which is shown in Figure 4. The low-frequency data below 1 kHz could not be recorded due to the low signal-to-noise ratio. The real ($\varepsilon'$) and imaginary ($\varepsilon''$) parts of the dielectric constant are presented in Figure 4a and 4b. The dielectric peak appears at the onset of magnetic ordering $T_{N2}$ as a result of strong magnetodielectric coupling, as reported earlier. Interestingly, the peak present in $\varepsilon'(T)$ and $\varepsilon''(T)$ exhibits a frequency-dependent behavior around $T_{N2}$. The peak temperature ($T_P$) of $\varepsilon'(T)$ and $\varepsilon''(T)$ shifts to a higher temperature with increasing frequency ($v$). Figure 4c shows the plot of $\ln(v)$ versus the inverse of $T_p$ (peak temperature extracted from $\varepsilon''$), showing it follows an Arrhenius relation ($v = v_0 \exp(E_a/k_B T_p)$, $v_0 =$ pre-exponential factor, $E_a =$ activation energy), exhibits logarithmic relaxation behavior with $E_a \sim 574$ K. The frequency dependence behavior can be observed due to the presence of dipolar glass behavior or ferroelectric domain relaxation/ reorientation. These results combined with ac susceptibility and neutron results (discussed below) indicate the presence of finite-size ferroelectric domains in this compound. The activation energy is related to the energy barriers associated with the ferroelectric domain reorientation. Very low polarization value in PUND measurement is consistent with the concept of finite-size ferroelectric (magnetoelectric) domains instead of true long-range ordering. A similar effect was reported in the well-known multiferroic compound $Ca_3CoMnO_6$ where a low polarization value was experimentally obtained due to the finite-size domains instead of true long-range ordering. [14]

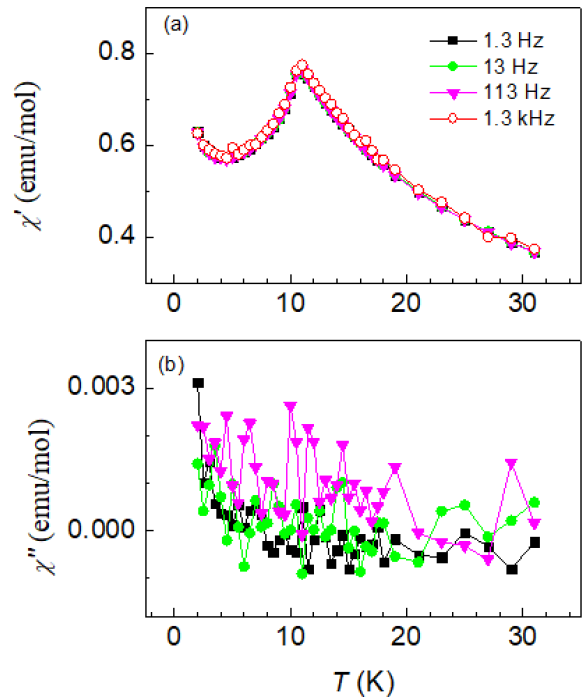

**Figure 5 (a) The real part ($\chi'$) and (b) imaginary part ($\chi''$) of ac magnetic susceptibility as a function of temperature for various frequencies from 1.3 Hz- 1.3 kHz.**

The ac magnetization as a function of temperature is shown in Figure 5 for different frequencies (1 Hz-1.3 kHz) which does not yield any frequency dependence behavior, in contrast with the dielectric constant. We propose this could be an artifact arising from a different frequency regime in which the finite-size magnetic domains respond to higher frequencies. Similar behavior is observed in the magnetism-driven ferroelectric system $Dy_2BaNiO_5$ where the frequency-dependent behavior is observed only in high-frequency regions (>10 kHz). [16] The response of the finite-size magnetic domains can be very weak for low-frequency regions.

We have analyzed the peak shape in neutron diffraction to clearly resolve this issue. We have superimposed three normalized representative peaks obtained from neutron diffraction in Figure 2b. We have considered the different peaks at T= 4 K; one is (1 0 1)-peak for the nuclear structure (only lattice Bragg peak), the other one is (-0.5 1 1) peak which is a magnetic peak related to $k_1$-spin structure ($k_1$= ½ 0 0), and the third one is (-0.75 0.25 1) magnetic peak for $k_2$-spin structure (magnetic Bragg peak related to $k_2$= ¼ ¼ 0). The (-0.5 1 1) magnetic peak related to the $k_1$-spin structure is well fitted with the (gaussian) peak shape parameters same as the (1 0 1) lattice peak used in the refinement (Figure 2b). The (-0.75 0.25 1)-peak shape associated with the $k_2$-spin structure shows significant broadening when compared to that of the $k_1$-spin structure (see Figure 2b). One needs to introduce significant Lorentzian parameters to fit the peaks associated with $k_2$-spin structure. Therefore, we only observe the broadening of the magnetic peak related to $k_2$-spin-strcture, no broadening is observed for the magnetic peak associated to $k_1$-spin structure even at low temperature below $T_{N2}$. This shows that the $k_2$-spin structure has a shorter coherence length compared to that of $k_1$-spin structure and it arises through the development of smaller magnetic domains. This results together with our frequency dependance magnetic and dielectric results predict the magnetoelectric domain dynamics.

These results clearly support our electric polarization model, where spin-driven ferroelectric polarization arises from the $k_2$-spin structure, however, the net polarizations (of a domain) nearly cancel out due to random domain dynamics of magnetoelectric domains in bulk macroscopic scale yielding very weak experimentally observed polarization. In this structure, the spins are oriented with ↑↑↓↓-spin structure of Ho and Ru-spins individually and non-colinear between Ho and Ru-spins. A higher magnetic field may destabilize the spin-structure, where spins will try to reorient along the applied H. Eventually, a spin can be flipped (↑↑↑↓) by changing the whole spin structure, depending on the strength of the magnetic field. The H-induced magnetic transition ~36 kOe and the M(T) feature at a very high field (50 kOe) are consistent with such behavior. [19] If there is a change of spin-pattern, then definitely the canting angle of Ru and Ho spins will be different, therefore, the direction of the vector $[\boldsymbol{e_{ij}} \times (\boldsymbol{S_i} \times \boldsymbol{S_j})]$ will be different. The relative spin-orientation between $S_{Ru}$ and $S_{Ho}$ under a high magnetic field is such that overall polarization may cancel out on a macroscopic scale. The shifting of the dielectric peak towards lower temperature with increasing magnetic field, and further the suppression (absence) of the dielectric feature under the high-magnetic field of 50 kOe [29] is consistent with this (spin-driven) ferroelectric polarization model.

## 3.3. Influence of External Pressure:

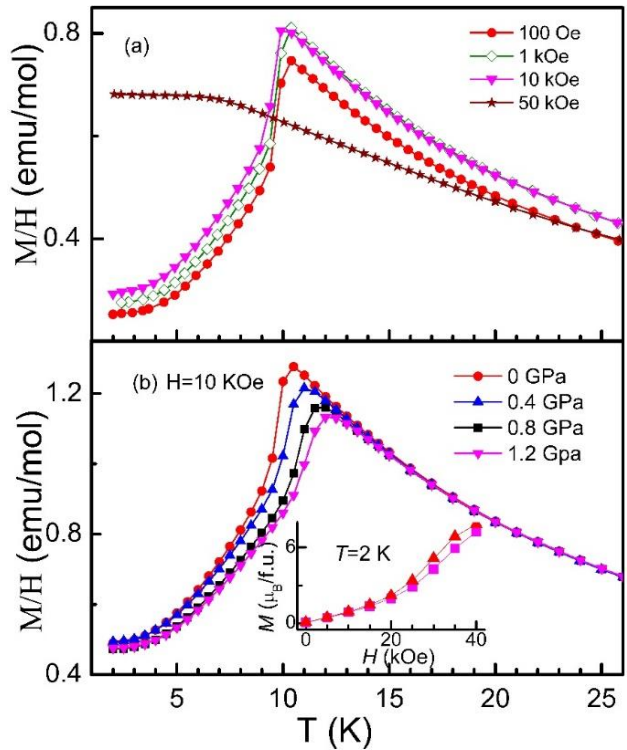

**Figure 6: DC magnetic susceptibility as a function of temperature, (a) in different applied field for zero applied pressure, and (b) under different applied pressures (0, 0.4, 0.8 and 1.2 GPa) in the presence of 10 kOe magnetic field. The inset of (b) shows the isothermal magnetization M(H) in the absence of pressure and the presence of 1.2 GPa pressure.**

The external parameters (e.g. field, pressure, etc.) have a strong effect on the magnetic (multiferroic) ordering of material. Earlier, we observed that the application of a magnetic field decreases the ordering temperature as observed from the shifting of T-dependence magnetic and dielectric peak towards low temperature with increasing magnetic field and a very high magnetic field (say, H=50 kOe) destroys the multiferroicity, [29] consistent with our spin-driven ferroelectric model. Here we investigate the effect of another parameter, that is, external pressure on the magnetic (multiferroic) ordering of this system. The dc magnetic susceptibility (M/H) under different magnetic fields is shown in Figure 6a. Temperature-dependent magnetization shows a peak at ~10.2 K as reported in the literature [29,35]. This peak shifts to a lower temperature and broadens with an increasing magnetic field, eventually being suppressed at 50 kOe, agreeing with the previous report. [29] This result confirms the reproducibility of the magnetic feature where the pressure cell does not have any extrinsic effect on the measurements. We have plotted the dc magnetization under external pressure at a fixed magnetic field of 10 kOe in Figure 6b. Interestingly, the temperature of the magnetic peak increases with increasing external pressure. The magnetic peak at 10.2 K in the absence of external pressure shifts to 11, 11.5 and 12.2 K under the application of small pressure of 0.5, 0.8 and 1.2 GPa, respectively, corresponding to an enhancement of ~ 1.6 K/GPa. These results suggest that external pressure and magnetic field have opposing effects on the magnetic ordering temperature. The inset of Figure 6b shows the isothermal magnetization plot as a function of the magnetic field for the absence and presence of external pressure of 1.2 GPa at T=2 K. We observe a small decrease in the magnetization value at 2 K in M(H) under external pressure which is consistent with M(T) behavior.

In the well-known spin-driven ferroelectric system, $DyMn_2O_5$, the application of pressure enhances the ferroelectric polarization at low temperatures by stabilizing the commensurate Mn-spin order.[24] In $RMnO_3$ (R = Dy, Tb, and Gd), the system was found to show a pressure-induced phase transition into the phase with large spin-driven ferroelectric polarization along 'a'- axis.[25] Density functional simulations suggest that the enhancement of polarization by applying pressure in $TbMnO_3$ is due to the stabilization of a collinear up-up-down-down spin-ordered state.[25] Here we speculate that similarly to previous reports on other systems, pressure (mildly) stabilizes the up-up-down-down structure resulting in an enhancement of magnetic ordering competing with temperature fluctuations.

As mentioned earlier, attempts were made to collect neutron diffraction under external pressure, but the weak neutron transmission through the cell body, along with aspects discussed in the earlier experimental section precluded the collection of usable data for structural and magnetic refinements.

### *3.4. Temperature dependence X-ray diffraction:*

We have analyzed the synchrotron high-resolution powder XRD data using Rietveld refinement [38] in the temperature range of 4-300 K. The Rietveld refinement fitting of HRXRD data at 8 K and 80 K are shown in Figure 7. We observed that the crystal symmetry remains consistent in temperature fitting with P63/mmc (194) space group. The T-dependence plot of lattice parameters ($a$, $c$) is represented in Figures 8a and 8b from 4-40 K. [39,40] A clear anomaly in the lattice parameters ($a$, $c$) around the transition temperature $T_{N2}$ is observed.

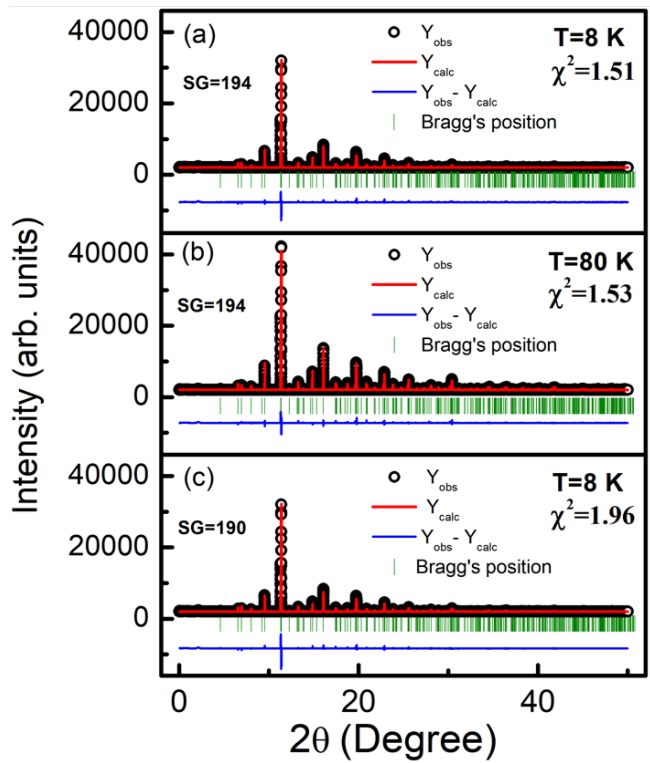

**Figure 7 Synchrotron powder X-ray diffraction pattern of $Ba_3HoRu_2O_9$ (a) at 8 K and (b) 80 K for space group (SG) 194, and (c) Synchrotron powder X-ray diffraction pattern of $Ba_3HoRu_2O_9$ at 8 K for space group 190 in Rietveld refinement.**

Cooling from the high temperature, the lattice constants ($a$, $c$) reduce with decreasing temperature due to thermal contraction. However, at the onset of the $T_{N2}$, the curve goes upward for lattice constant 'c' with the decreasing temperature, which is not usual. The lattice constant '*a*' also shows a peak at $T_{N2}$. The change in the lattice volume (V) with temperature is shown in Figure 8 (c) which displays a clear anomaly around $T_{N2}$ in the form of a peak. Considering that lattice volume is directly linked to the Gibbs free energy (G), $V = \left(\frac{\partial G}{\partial P}\right)_T$, any discontinuity in volume implies a discontinuity in Gibbs free energy. This observation indicates a possible first-order type phase transition. [41] A sharp change in T-dependence heat capacity is observed at $T_{N2}$ predicting the first-order nature of the phase transition [27], consistent with this result.

We have also checked some other non-centrosymmetric space groups related to this structure (where spatial inversion symmetry is broken), though, we do not find any further improvement in the goodness of fit with these related space groups. We have analyzed the XRD data with a non-centro symmetric space group with a similar structure. The Rietveld refinement fitting at T=8 K for the non-centro symmetric space group, $P\bar{6}2c$, is shown in Figure 7c. We found equally good fitting for both the space groups at 8 K. It is to be noted that the displacement of atoms in multiferroic-II (spin-driven ferroelectric) is generally very small to be traced in the X-ray/ neutron diffraction. Till now, for all the reported spin-driven ferroelectric systems, the breaking of spatial inversion symmetry is not traceable directly from the XRD analysis, rather an anomaly in the lattice parameters around multiferroic ordering is reported in the literature. [27,41,42] Friedel's law states that the intensities of the h, k, l and -h, -k, -l reflections are equal. This is true if the crystal is centrosymmetric or if no resonant scattering is present.

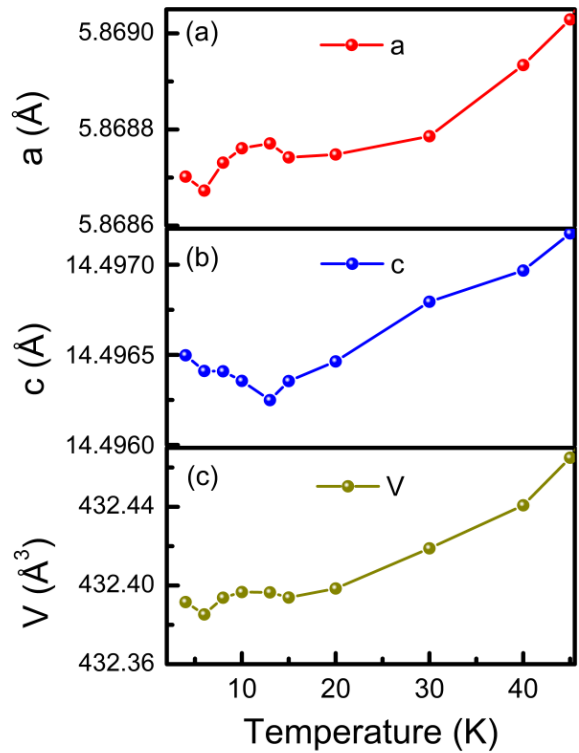

**Figure 8: (a) and (b) Thermal variation of lattice parameters (*a*- and *c-axis*) respectively. (c) thermal variation of lattice volume. The symbol size represents the error of the data.**

It is impossible to conclude by X-ray powder diffraction whether an inversion center is present unless significant structural change is there in general. Combining the results of synchrotron X-ray and bulk ferroelectric measurements, we propose the non-centrosymmetric space group P$\bar{6}$2c for this title compound.

## 4. Conclusion:

We have investigated in detail the mechanism of spin-driven electric polarization on $Ba_3HoRu_2O_9$, an ideal model for a $4d$-$4f$ coupled system, through dielectric spectroscopy, magnetization measurements, time-of-flight neutron diffraction, and temperature-dependent synchrotron X-ray diffraction. The ferroelectricity below $T_{N2}$ arises from inverse D-M interaction between the non-collinear moments of Ho and Ru-magnetic ions of 2$^{nd}$ magnetic phase related to $k_2$-spin structure, where each Ho and Ru-spins form a collinear ↑↑↓↓ spin-structure but Ho and Ru ions are non-collinear connected via Ru-O-Ho exchange path. This is, to the best of our knowledge, the first experimental demonstration of spin-driven ferroelectricity in a $4d$-$4f$ systems, not only that, spin-driven polarization, arising due to inverse D-M interaction of two different magnetic ions, (here, Ho($4f$) and Ru($4d$)), is rarely (experimentally) demonstrated in any $d$-$f$ system. Only for the well-known multiferroic $GdMn_2O_5$ compound (which is considered to be one of the best spin-driven ferroelectric compounds in which the largest polarization is observed among multiferroic-II systems), it is demonstrated that the exchange-striction mechanism between collinear spins of Gd($4f$) and Mn($3d$) take parts in contributing to larger polarization. The weak polarization observed in our titled compound $Ba_3HoRu_2O_9$ is attributed to the partial cancellation of magnetoelectric domains, as estimated from the excess broadening of on 2$^{nd}$ magnetic phase ($k_2$- spin structure) in the neutron diffraction. The clear anomaly in the lattice parameters and volume obtained from high-resolution synchrotron XRD results also supports the ferroelectric phase transition at $T_{N2}$. Our synchrotron XRD results propose a new non-centrosymmetric space group (P$\bar{6}$2c) for this compound. An application of a very high magnetic field destroys such spin-structure and thus ferroelectricity. Interestingly, the application of external pressure has an opposite effect on the magnetic ordering ($T_{N2}$) compared to the external magnetic field. The application 1.2 GPa pressure enhances the magnetic ordering $T_{N2}$ by 2 K. Other microscopic experiments such as µSR spectroscopy, polarized neutron diffraction [43] under extreme condition are warranted for further understanding the spin dynamics of Ba$_3$HoRu$_2$O$_9$. Our results hopefully motivate further investigations on new 4*d*-4*f* coupled systems where the magnetic phase might be stabilized and may yield a large polarization.

## Acknowledgements:

TB greatly acknowledges the Science and Engineering Research Board (SERB) (Project No.: SRG/2022/000044), and UGC-DAE Consortium for Scientific Research (CSR) (Project No CRS/2021-22/03/544), Government of India for funding. DTA would like to thank EPSRC-UK (Grant No. EP/W00562X/1) for funding. A portion of this research used resources at the Spallation Neutron Source (neutron diffraction), and the Center for Nanophase Materials Sciences (High-pressure magnetization), both DOE Office of Science User Facilities operated by the Oak Ridge National Laboratory, USA.

We thank Erik Elkaim, Synchrotron SOLEIL, L'Orme des Merisiers,Saint-Aubin, BP48, 91192 Gif sur Yvette, France for his help in performing synchrotron XRD experiment at CRISTAL beamline in SOLEIL synchrotron(France). We are also thankful to Dr. Somnath Ghara, University of Augsburg, Augsburg, Germany; and Dr. S. D. Kaushik, UGC–DAE Consortium for Scientific Research, Mumbai, India for fruitful discussion.

The authors declare no competing financial interest.

# Supplementary Information

## Calculation of Inverse D-M interaction related to $K_2$-spin structure:

Fig.S1 shows the detailed crystal and $K_2$-spin structure of the title compound. The Ho and Ru spins are connected via the Ho-O-Ru-O-Ho exchange path. The individual Ho-spins and individual Ru-spins are colinear to each other (↑↑↓↓). However, the neighbouring Ru and Ho spins are directly connected via oxygen and not colinear to each other. Therefore, these non-colinear Ru and Ho spins can create non-zero polarization via inverse D-M interaction. There are four different configurations of this non-linear zig-zag spins of Ru and Ho which are periodical. For better understanding, we have shown these configurations in Fig. S2 for ac-pane and ab-plane respectively.

We have calculated the inverse D-M interaction $[\boldsymbol{P_{ij}} \propto e_{ij} \times (\boldsymbol{S_i} \times \boldsymbol{S_j})]$ of the neighbouring Ho and Ru-spins for each configuration in the magnetic unit cell as shown in Fig S1 and S2. The detailed calculation for Configuration -1 is described below, which creates a non-zero local polarization. In the same fashion, we have calculated the $[P_{ij} \propto e_{ij} \times (S_i \times S_j)]$ for other configurations as well. Interestingly, the overall polarization in this magnetic cell does not get cancelled out.

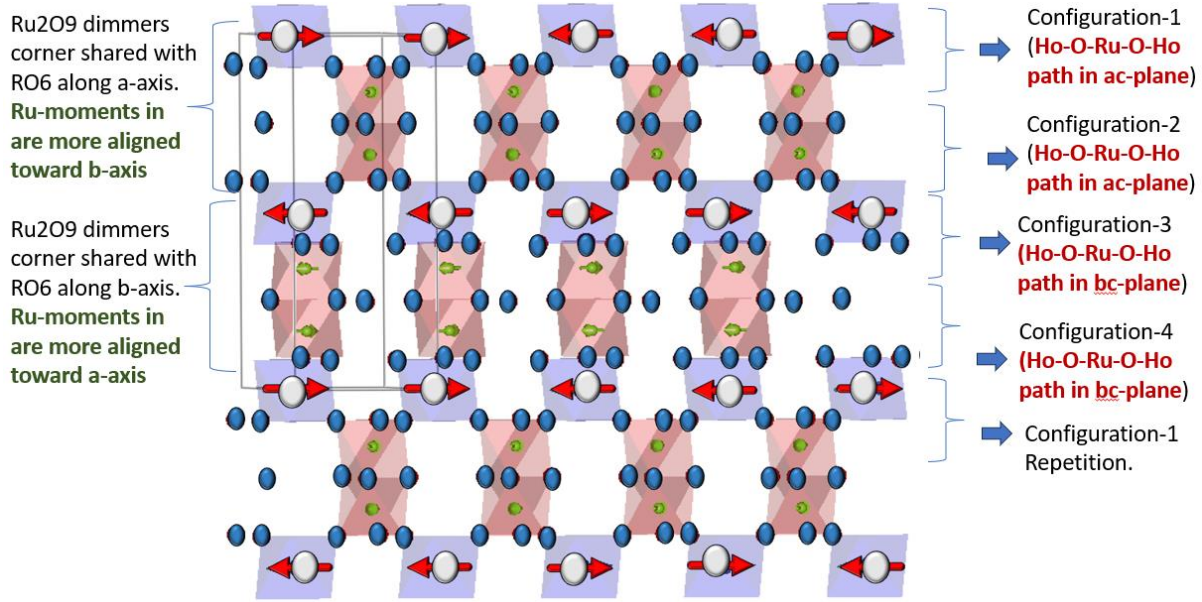

**Fig.S1 Lattice structure along with spin-patterns related to K2-spin structure of $Ba_3HoRu_2O_9$ in ac-plane.**

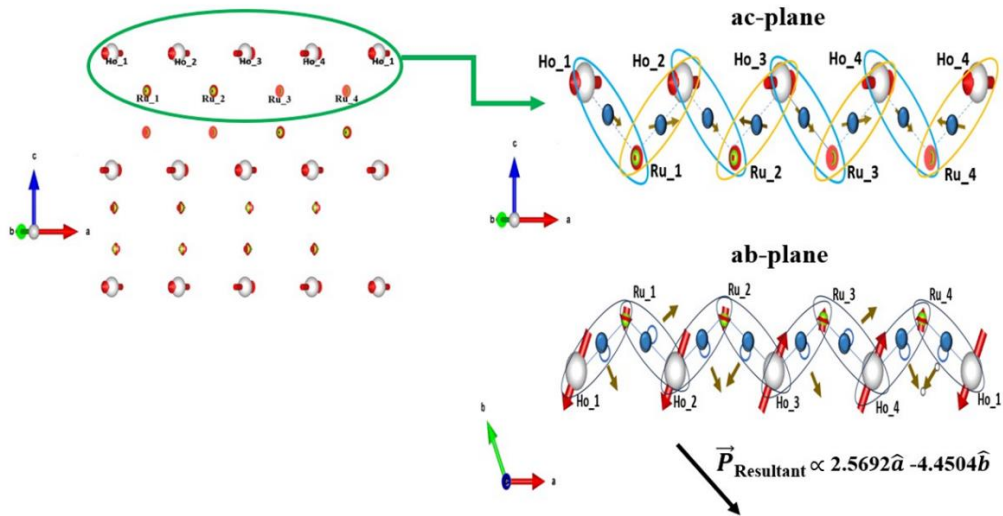

**Fig.S2: (Left) The spin-pattern related to the K2-spin structure of $Ba_3HoRu_2O_9$ in ac-plane (same as Fig.S1) omitting oxygen atoms. (Right) The inverse D-M interaction between Ru and Ho spins mediated via intermediate O atoms are shown for one zig-zag layer (Configuration-1 in Fig. S1). For clarity this is shown both in ac-plane and the ab-plane. The White, Green and Blue denotes the Ho, Ru and Oxygen atoms respectively, the Red arrow denotes the spins moments of Ho and Ru on these ions and brown arrow denotes the direction of inverse D-M interaction act between these two spins. The resultant Inverse D-M interaction (that is, resultant polarization) is shown with large black arrow.**

We have, $\vec{P}_{(Ho1-Ru1)} \propto \vec{e}_{(Ho1Ru1)} \times (\vec{S}_{Ho1} \times \vec{S}_{Ru1})$, where $\vec{e}_{(Ho1Ru1)}$ is the direction between Ho1 and Ru1 (unit vector), $\vec{S}_{Ho1}$ and $\vec{S}_{Ru1}$ are spins of Ho1 and Ru1 respectively, as shown in Fig.S2.

Here, $\vec{R}_{(Ho1Ru1)} = \vec{R}_{Ru1} - \vec{R}_{Ho1}$

$\vec{R}_{(Ho1Ru1)} = 2.9270\hat{a} + 1.6898\hat{b} - 2.3370\hat{c}$

$\vec{e}_{(Ho1Ru1)} = 0.7123\hat{a} + 0.4112\hat{b} - 0.5687\hat{c}$

$(\vec{S}_{Ho1} \times \vec{S}_{Ru11}) = (-4.11528\hat{a} - 4.11528\hat{b} + 0\hat{c}) \times (-1.24506\hat{a} - 0.86552\hat{b} + 0\hat{c})$

$(\vec{S}_{Ho1} \times \vec{S}_{Ru1}) = 0\hat{a} + 0\hat{b} - 1.5619133712\hat{c}$

Therefore

$\vec{P}_{(Ho1-Ru1)} \propto 0.6423\hat{a} - 1.1126\hat{b} + 0\hat{c}$

All these values of positions and spins of Ho and Ru-ions are taken from magnetic structure as obtained from neutron diffraction.

Similarly, we have calculated the polarization for all other configurations in the magnetic cell.

$\vec{P}_{(Ru1-Ho2)} \propto (0.6423\hat{a} + 1.1126\hat{b} + 0\hat{c})$

$\vec{P}_{(Ho2-Ru2)} \propto (0.6423\hat{a} - 1.1126\hat{b} + 0\hat{c})$

$\vec{P}_{(Ru2-Ho3)} \propto (-0.6423\hat{a} - 1.1126\hat{b} + 0\hat{c})$

$\vec{P}_{(Ho3-Ru3)} \propto (0.6423\hat{a} - 1.1126\hat{b} + 0\hat{c})$

$\vec{P}_{(Ru3-Ho4)} \propto (0.6423\hat{a} + 1.1126\hat{b} + 0\hat{c})$

$\vec{P}_{(Ho4-Ru4)} \propto (0.6423\hat{a} - 1.1126\hat{b} + 0\hat{c})$

$\vec{P}_{(Ru4-Ho1)} \propto (-0.6423\hat{a} - 1.1126\hat{b} + 0\hat{c})$

Total (resultant) polarisation,

$\vec{P}_{Resultant} \propto (\vec{P}_{(Ho1-Ru1)} + \vec{P}_{(Ru1-Ho2)} + \vec{P}_{(Ho2-Ru2)} + \vec{P}_{(Ru2-Ho3)} + \vec{P}_{(Ho3-Ru3)} + \vec{P}_{(Ru3-Ho4)} + \vec{P}_{(Ho4-Ru4)} + \vec{P}_{(Ru4-Ho1)})$

$\vec{P}_{Resultant} \propto [0.6423\hat{a} - 1.1126\hat{b} + 0\hat{c} + 0.6423\hat{a} + 1.1126\hat{b} + 0\hat{c} + 0.6423\hat{a} - 1.1126\hat{b} + 0\hat{c} - 0.6423\hat{a} - 1.1126\hat{b} + 0\hat{c} + 0.6423\hat{a} - 1.1126\hat{b} + 0\hat{c} + 0.6423\hat{a} + 1.1126\hat{b} + 0\hat{c} + 0.6423\hat{a} - 1.1126\hat{b} + 0\hat{c} - 0.6423\hat{a} - 1.1126\hat{b} + 0\hat{c}]$

$\vec{P}_{Resultant} \propto (2.5692\hat{a} - 4.4504\hat{b})$